\documentclass[12pt]{article}
\usepackage{color}
\definecolor{myblue}{rgb}{.8, .8, 1}

\usepackage{amsmath} 
\usepackage{empheq}
\usepackage[bbgreekl]{mathbbol}

\newlength\mytemplen
\newsavebox\mytempbox

\makeatletter
\newcommand\mybluebox{%
    \@ifnextchar[%]
       {\@mybluebox}%
       {\@mybluebox[0pt]}}

\def\@mybluebox[#1]{%
    \@ifnextchar[%]
       {\@@mybluebox[#1]}%
       {\@@mybluebox[#1][0pt]}}

\def\@@mybluebox[#1][#2]#3{
    \sbox\mytempbox{#3}%
    \mytemplen\ht\mytempbox
    \advance\mytemplen #1\relax
    \ht\mytempbox\mytemplen
    \mytemplen\dp\mytempbox
    \advance\mytemplen #2\relax
    \dp\mytempbox\mytemplen
    \colorbox{myblue}{\hspace{1em}\usebox{\mytempbox}\hspace{1em}}}

\makeatother

\usepackage{bbm}
\usepackage{yfonts}
\pdfoutput=1
\usepackage{amsfonts}
\usepackage[bbgreekl]{mathbbol}
\makeatletter
\def\amsbb{\use@mathgroup \M@U \symAMSb}
\makeatother
\usepackage[bbgreekl]{mathbbol}
\usepackage{amssymb,amsfonts}
\usepackage{graphicx}
\usepackage[top=2.75cm, bottom=2.75cm, left=2.75cm, right=2.75cm]{geometry} 
\usepackage{appendix}
\usepackage{color}

\newcommand{\be}{\begin{equation}}
\newcommand{\ee}{\end{equation}}
\newcommand{\ber}{\begin{eqnarray}}
\newcommand{\eer}{\end{eqnarray}}

\def\+{{+\!\!\!+}}

\newcommand{\pa}{\partial}
\newcommand{\na}{\nabla}

\newcommand{\bbD}[1]{\mathbb{D}_{#1}}

\newcommand{\h}{\theta}
\newcommand{\then}{\Rightarrow}
\newcommand{\al}{\alpha}
\newcommand{\La}{\Lambda}
\newcommand{\bh}{\bar\theta}
\newcommand{\ald}{{\dot\alpha}}

\newcommand{\eps}{\varepsilon}

\begin{document}
\begin{titlepage}
\begin{flushright} \small
%%% Preprint numbers%%%
UUITP-13/25\\
YITP-SB-2025-09\\
\end{flushright}
%%%%%%%%%%%
\smallskip
\begin{center} 
\LARGE
{\bf  Superspace Supergravity} \\[20mm] 
\large
{\bf Ulf~Lindstr\"om$^a$} and {\bf Martin Ro\v cek$^{b}$} \\[20mm]
{ \small\it
${}^a$Department of Physics and Astronomy, Division of Theoretical Physics,
Uppsala University, \\ Box 516, SE-75120 Uppsala, Sweden \\ and \\
Center for Geometry and Physics, Uppsala University, Box 480, SE-75106 Uppsala, Sweden\\~\\
${}^b$C.N.~Yang Institute for Theoretical Physics, Stony Brook University, Stony Brook NY 11794-3840, USA}

\end{center}

\vspace{10mm}
\centerline{\bfseries Abstract} 
We give a brief, incomplete, and idiosyncratic review of the early years of supergravity in superspace 
as our contribution to the book {\em Half a Century of Supergravity} edited by Anna Ceresole and Gianguido Dall'Agata.
\bigskip

\end{titlepage}

\section{Introduction}
In this brief contribution, we discuss aspects of 4 dimensional superfield supergravity; our presentation is largely complementary to the contribution by S.J. Gates. 

Of course, supergravity first constructed in components by Ferrara, Freedman, and van Nieuwenhuizen \cite{Freedman:1976xh}\footnote{The contribution to this volume of P. van Nieuwenhuizen has a discussion of superspace supergravity in quantum mechanics.}, but rather quickly, the superspace formulation was developed; this is the focus of our brief contribution.

Superspace for rigid supersymmetry was introduced by Salam and Strathdee \cite{Salam:1974yz} and a few months later by Ferrara, Wess, and Zumino \cite{Ferrara:1974ac}, and involved appending anticommuting spinor coordinates $\h$ to the usual coordinates $x$ of spacetime. We review the description of super-Yang-Mills theory from \cite{Ferrara:1974pu} (with a somewhat more modern perspective and using different conventions). In 4 dimensions, $N=1$ superspace has coordinates $x^m, \h^\al,\bh^\ald$. Superfields $\Phi(x,\h,\bh)$ are fields that depend on these coordinates; new superfields can be made by differentiation with covariant spinor derivatives
\be
D_\al=\frac{\pa}{\pa\h^\al} +\frac{i}2\bh^\ald\pa_{\al\ald}~,~~
\bar D_\ald=\frac{\pa}{\pa\bh^\ald} +\frac{i}2\h^\al\pa_{\al\ald}~,
\ee
where $\pa_{\al\ald}$ is the usual $x^m$ derivative written as a $2\times2$ matrix. These obey the superalgebra
\be
\{D_\al,D_\beta\}=0~,~~ \{D_\al,\bar D_\ald\} =i\pa_{\al\ald}~,~~[D_\beta,\pa_{\al\ald}]=0~,~etc.
\ee
Matter is typically described by complex chiral scalar superfields obeying
\be
\bar D_\ald \Phi=0~,~~ D_\al \bar\Phi = 0~,~~ \bar\Phi\equiv\Phi^\dag
\ee
Component fields can be found either by explicit Taylor expansions in $\h^\al,\bh^\ald$ or by covariant expansions using $D_\al,\bar D_\ald$.

In general, $\Phi$ can carry a representation of a gauge group. Since the superfields are chiral, it is natural to consider local gauge transformations with chiral parameters:
\be
\Phi\to e^{i\La}\Phi~,~~ \bar\Phi\to \bar\Phi e^{-i\bar\La}~.
\ee
The gauge field is described by a Hermitian scalar $V$ which can be used to construct gauge-covariant derivatives and gauge invariant Lagrangians; in so-called chiral representation\footnote{Since the parameter $\La$ is complex, this representation is not real; in particular, $\bar\na_\ald\ne(\na_\al)^\dag$.}, $V$ can be used to construct objects that only transform under $\La$ and not under $\bar\La$:
\be
\bar\na_\ald=\bar D_\ald~,~~\na_\al=e^{-V}D_\al e^V~,~~ \bar\Phi e^V~,
\ee
where 
\be
e^V\to e^{i\bar\La} e^V e^{-i\La}~~~~\then~~~\mathcal{L}=\bar\Phi e^V \Phi ~~{\rm~is ~invariant.}
\ee
The gauge-covariant derivatives obey a modified superalgebra with some ({\em but not all curvatures}) non-zero:
\be
\{\na_\al,\na_\beta\}=0~,~~ \{\na_\al,\bar \na_\ald\} =i\na_{\al\ald}~,~~
[\na_\beta,\na_{\al\ald}]=\eps_{\al\beta}\bar W_\ald~,~etc.
\ee
In addition to some curvatures vanishing, the remaining non-zero ones obey constraints, e.g.,
$\na_\beta\bar W_\ald=0$.
Again, component fields can be found either by explicit Taylor expansions in $\h^\al,\bh^\ald$ or by covariant expansions using $\na_\al,\bar \na_\ald$.

The first attempts to describe supergravity in superspace were made by Arnowitt and Nath \cite{Nath:1975nj} by naively extending General Relativity to superspace; this theory was not unitary and suffered from a large number of extra higher spin fields. 

Wess and Zumino realized that since the tangent space of superspace is a reducible representation of the Lorentz group, namely spinor $\oplus$ vector, it was natural to consider different irreducible components of the curvature and torsion tensors, and look for consistent constraints on them--indeed, we already saw that the Yang-Mills supercurvatures obey such constraints. In the notation of \cite{Gates:1983nr}, the torsions and curvatures are defined as:
\be
[\na_A,\na_B\}=T_{AB}{}^C\na_C+R_{AB}(M)~.
\ee
In \cite{Wess:1977fn}, Wess and Zumino proposed superspace constraints corresponding to field equations; in \cite{Grimm:1977kp} they proposed off-shell constraints, and in \cite{Wess:1978bu} they found an action whose linearized variation gave the field equations.  Up to some redefinitions, their constraints are equivalent to those of \cite{Gates:1983nr}:
\be
T_{\al,\beta}{}^{\dot\gamma}=T_{\al,\beta}{}^{\gamma\dot\gamma}=T_{\al,\beta\dot\beta}{}^{\gamma\dot\gamma}=R_{\al,\dot\beta\,\gamma\delta}=R_{\al,\beta\,\dot\gamma\dot\delta}=0~,~~
T_{\al,\dot\beta}{}^{\gamma\dot\gamma}=i\delta_\al{}^\gamma \delta_{\dot\beta}{}^{\dot\gamma}~,
\ee
Wess and Zumino did not solve their constraints in terms of unconstrained superfields.

The constraints were solved in a remarkable series of preprints by Warren Siegel, who took the perspective that superdiffeomorphisms should be treated as the gauge group of the translations in superspace, and basically copied the approach of super-Yang-Mills, with the Lie-algebra valued superfield $V$ replaced by $iU^M\pa_M$ \cite{Siegel:1977ab,Siegel:1977ng}\footnote{These preprints can be downloaded from INSPIRE.}. Complications arise because the group is not unimodular, but he was able to construct the action in terms of unconstrained superfields and thus solve the Wess-Zumino torsion constraints. Because the community was unfamiliar with his work and his approach, these seminal articles were never published; a beautiful pedagogical paper based on them, coauthored with S.J. Gates\cite{Siegel:1978mj}, was published and became very influentiual. It explained and greatly expanded the construction. Among the results in \cite{Siegel:1977ab,Siegel:1977ng,Siegel:1978mj} is the understanding of so-called conformal compensators that give rise to different component auxiliary fields and modified torsion and curvature constraints. 

The nonlinear relation of superspace supergravity to component supergravity using the explicit Taylor series approach was given in \cite{Rocek:1978yz,Rocek:1978hk}. The basic idea was to express the Taylor coefficients of superfields in superspace in terms of known component fields; 
Wess and Zumino gave an alternate approach using a covariant superderivative expansion in \cite{Wess:1978ns}.\footnote{A related procedure called ``Gauge Completion'' can be used to go from component theories to superspace \cite{Nath:1976ci,Ferrara:1979pk}.}

Whereas in Einstein's theory, the Bianchi identities are simply identities, in superspace supergravity, since the tangent space is reducible and we impose constraints on some torsions and curvatures, they give nontrivial constraints on other components and led to a number of works ``solving the Bianchi identities"  \cite{Grimm:1978ch,Howe:1981gz}.

All these developments gave rise to a variety of applications: the unconstrained description of superspace supergravity led to powerful superspace Feynman graph techniques for the theory \cite{Grisaru:1981xm, Grisaru:1982zh}. The components of superspace gave a convenient way to understand the superHiggs effect in superspace using nonlinear constrained superfields coupled to supergravity \cite{Lindstrom:1979kq}. 

Eventually, these and other developments found their way into a several books: \cite{Gates:1983nr,Wess:1992cp,Buchbinder:1998twe}.\footnote{These are the current active references on INSPIRE; the first edition of Wess and Bagger appeared in 1983, and of Buchbinder and Kuzenko in 1995.}

\section{Extended supergravity in $N=1$ superspace}
We now turn to some less well known applications of superspace supergravity--which perhaps can suggest some new research directions.

It has long been known that theories with more rigid supersymmetries can be described in flat $N=1$ superspace. In a series of little-known papers \cite{Gates:1980mf,Gates:1983ie,Gates:1984wc,Labastida:1984us,Labastida:1984qa,Labastida:1986md}, it was shown that this can be done for $N=2$ supergravity as well; it would be worthwhile to explore higher $N$ in this formulation. 
One benefit of this construction is that it gives the explicit solution to the torsion constraints in terms of just a few  unconstrained $N=1$ superspace prepotentials.

The basic idea is to use the approach of \cite{Wess:1978ns} to reduce the $N=2$ supergravity torsion and curvature constraints to $N=1$ superspace. $N=2$ superspace has twice as many fermionic $\h$ coordinates as $N=1$ superspace. We covariantly Taylor expand both fields and superdiffeomorphisms using the torsion and curvature constraints and choosing convenient gauges to obtain $N=1$ objects. Explicitly, we define $N=1$ superspace derivatives $\na_A$ by projecting some $N=2$ derivatives:
\be
\bbD{A}|=\na_A+\psi_A^\mu\pa_{2\mu}+\psi_A^{\dot\mu}\bar\pa_{\dot\mu}{}^2~,
\ee
where $\bbD{A}|$ is the $\h^{2\mu}$-independent projection of the $N=2$ superderivatives 
$\bbD{\al\ald},\bbD{1\al},\bbD{\ald}{}^1$. Note the apppearence of the new $N=1$ superfields $\psi_A^\mu,\psi_A^{\dot\mu}$; these describe the second gravitino of the the theory. 
Similarly, using the constraints, one can evaluate $\bbD{2\beta}D_A$, etc., and reduce all the $N=2$ fields to $N=1$. We also find the supersymmetry transformations of the $N=1$ superfields under the nonmanifest supersymmetry; evidently,
\be
\delta\psi_M{}^\al = \na_M\eps^\al+...
\ee
where $\eps$ is the parameter of the {\em nonmanifest} supersymmetry transformation.
The constraints are solved in terms of the $N=1$ superspace supergravity fields and an $N=1$ vector multiplet $V$ which is the gauge field of a central charge multiplet, a spinor prepotential $\varphi^\al$ for the second gravitino, and a chiral compensator $\Phi$.
\\~\\
{\bf Acknowledgement}\\
We thank Peter van Nieuwenhuizen for encouragement and comments, and Anna Ceresole for allowing us to submit our contribution long after the deadline. UL would like to thank the SCGP for their support while writing this contribution.

\eject

\end{document}